# USING A BAG OF WORDS FOR AUTOMATIC MEDICAL IMAGE ANNOTATION WITH A LATENT SEMANTIC

Riadh Bouslimi[1], Abir Messaoudi[2] and Jalel Akaichi[3]

Department of Computer Sciences
High Institute of Management, BESTMOD Laboratory
Tunis, Tunisia
[1]`bouslimi.riadh@gmail.com,`[2]`abir.messaoudi@gmail.com and`
[3]`jalel.akaichi@isg.rnu.tn`

## ABSTRACT

*We present in this paper a new approach for the automatic annotation of medical images, using the approach of "bag-of-words" to represent the visual content of the medical image combined with text descriptors based approach tf.idf and reduced by latent semantic to extract the co-occurrence between terms and visual terms. A medical report is composed of a text describing a medical image. First, we are interested to index the text and extract all relevant terms using a thesaurus containing MeSH medical concepts. In a second phase, the medical image is indexed while recovering areas of interest which are invariant to change in scale, light and tilt. To annotate a new medical image, we use the approach of "bag-of-words" to recover the feature vector. Indeed, we use the vector space model to retrieve similar medical image from the database training. The calculation of the relevance value of an image to the query image is based on the cosine function. We conclude with an experiment carried out on five types of radiological imaging to evaluate the performance of our system of medical annotation. The results showed that our approach works better with more images from the radiology of the skull.*

## 1. INTRODUCTION

In the last decade, a large number of medical reports containing textual information and digital medical images have been produced in hospitals. We distinguish several types of medical images include X-ray associated with medical reports such as: *magnetic resonance imaging* (MRI), *computed tomography* (CT), *magnetic source imaging* (MSI), *magnetic resonance spectroscopy* (MRS), etc. These medical images are stored in large databases. Further, images will be accessible to doctors, professionals, researchers and students to diagnose current patients and to provide valuable information for their studies.

Owing to the daily use of digital medical images, it is very important to develop techniques for information retrieval, which may improve the search efficiency in large databases of medical images. Among various advanced information retrieval, image annotation is considered an essential task for the management of image database [1]. If the images are annotated manually by

keywords which present an expensive and subjective task since it did not describe the content of the image and it provides many errors. Thus, manual annotation suffers from the following limitations, especially in image databases mass [2]. To remedy these limitations, the automatic annotation of images is necessary to make efficient image retrieval. Consequently, Automatic annotation of images is a hot topic in the field of multimedia information retrieval and machine learning. The automatic annotation is essentially based on the combination of textual and visual information.

The largest number of systems which deals with multimedia medical reports exploits only the textual part of these reports to extract the keywords and attribute them to the medical image. The standard approach is to represent the text as a "*bag-of-words*" [3] and to associate a weight for each word to characterize its frequency by the method *tf.idf*. The challenge today is to extend the system to other conditions, including the images because they are very active in multimedia collections. A natural approach is to use the same representation based on "bag-of-words" model to the image. This approach has already shown its effectiveness including applications for image annotation or search for objects in large collections [4].

The main goal of this article is to combine textual and visual information and to build a feature vector that is reduced by the latent semantic. If one has in entering a new un-annotated medical image, our system charge to find the suitable keywords that will be associated. In Section 2, we present the state of the art containing different existing works related on the combination of both textual and visual information, and automatic annotation of medical images. Then in Section 3, we present a first contribution is the system of automatic indexation of medical reports which combines both textual and visual information using the approach of "*bag-of-words*", we added in this system the *MeSH* thesaurus to control the indexed vocabulary. The combination of the two information will be reduced using the latent semantic to build co-occurrences between words and regions of interest in medical image. A second contribution presents an automatic annotation of a new medical image which consists in comparing the two histograms using measure of intersection. Finally, we conclude and present some perspectives.

## 2. REPRESENTATION OF IMAGES WITH A BAG OF VISUAL WORDS

Finding objects in images and matching images are generally carried directly using local features computed on the images [5]. In broader contexts such as the categorization of images or information retrieval, these local features are grouped using an algorithm for unsupervised classification to form a vocabulary of visual words as illustrated in Figure 1 [6, 7, 8]. The most common approach is dynamic cloud or K-means algorithm [9, 10]. The classification step patterns can be seen as a step of adaptive sampling of the feature space. This makes it possible to reduce the size of this space and then to calculate histograms of occurrence of visual words. A visual word is interpreted as a class of patterns present frequently in the images.

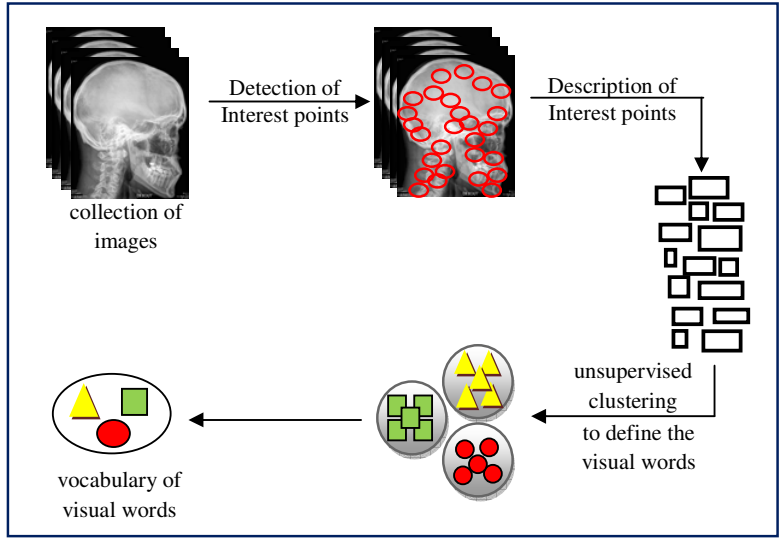

Figure1.creation of a vocabulary of visual words

The vocabulary of visual words is then used to represent the image as illustrated in Figure 2. The regions of the image are associated with visual words according to their descriptions. The image can then be represented through using a weighted vector of visual words. This representation that uses the model of visual bag of words depends on many parameters such as the choice of the detection and description of interest points [11], the choice of the classification algorithm [7], the size of the visual vocabulary, the number of visual words by calculating image and normalization [12]. In the following, these parameters will be a more precise study.

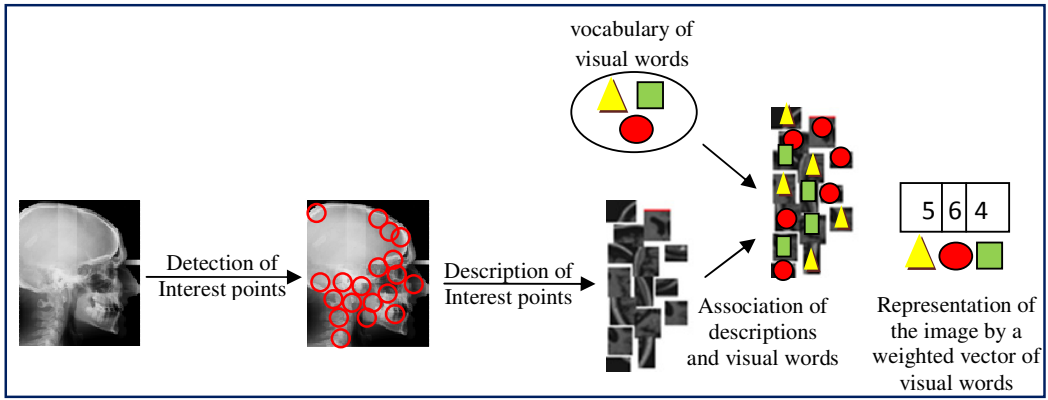

Figure2.Representation of the image using a bag of visual words model

Recent studies in images categorization showed that the detection of regular points of interest gives the best results [7], especially when the vocabulary size is important [12]. As for the representation of textual documents, visual words are weighted for each image. The occurrence of visual words is used, but the discriminating power of these is also taken into account [8].Some approaches use a binary weighting of visual words by considering the occurrence or non-occurrence of words in images [12]. A selection of visual words to consider representation is sometimes done by retaining only those that maximize the mutual information between a word and a class [12]. One of the current challenges concerning the description of visual information is taken into account the spatial information. Several approaches have been proposed in this area, based for example on a regular cutting of the image from a pyramidal decomposition or by constructing a visual phrase [13, 14, 15].

The choice of the representation of images is difficult, mainly because of the existence of the semantic gap between the description and the interpretation of these images. The approach is more and more used, described the images as bags of words are textual documents. This textual and visual information were presented separately, but to describe multimedia documents efficiently, it is essential to combine them.

## 3. INDEXING MEDICAL REPORTS

The medical report is proposed to describe the text and images with textual and visual terms. The two *modalities* are initially processed separately using each approach "*bag-of-words*". They are then represented as a vector of *tf.idf* weight characterizing the frequency of each word visual or textual. Use the same type of representation for the two modalities can be combined by a late fusion methodology and then perform queries to retrieve multimedia information. This general methodology is presented in Figure 3.

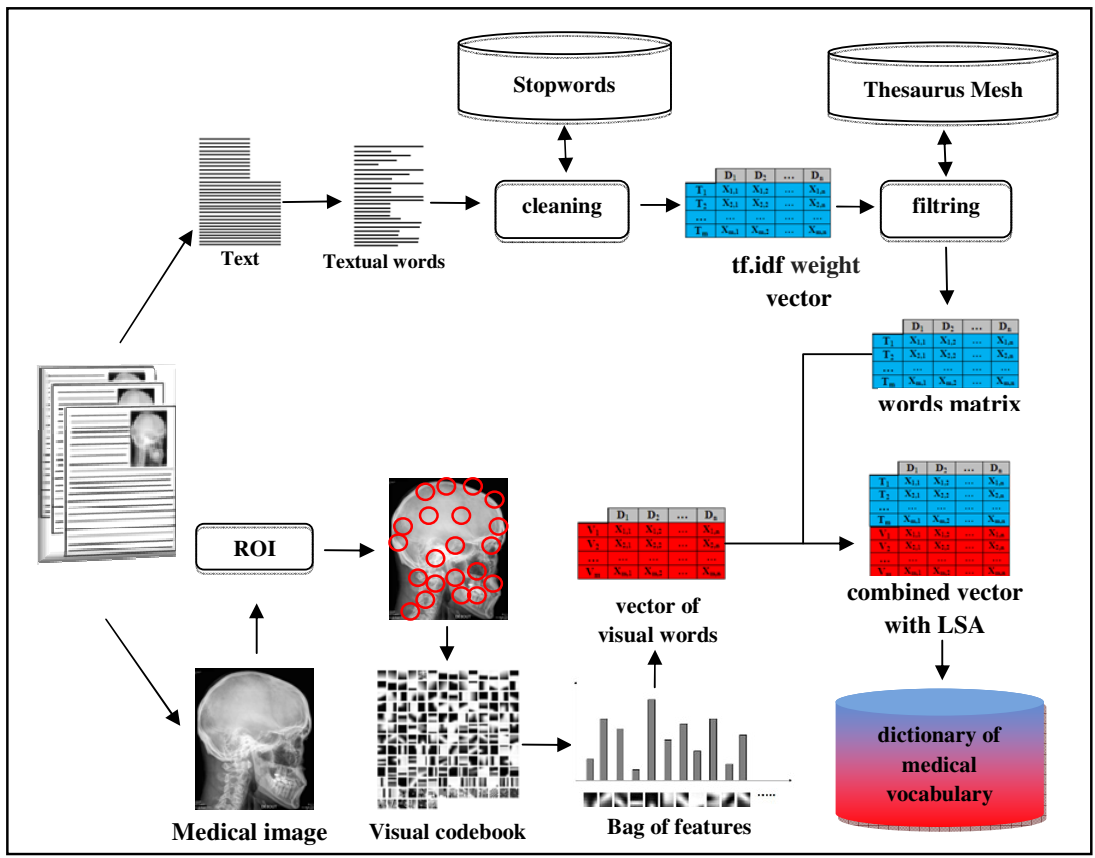

Figure3.Indexing medical report

### 3.1. Textual indexation of medical reports

To represent a text document as a vector of weights, first it is necessary to define a text index terms or vocabulary. To do this, we first apply a lemmatization Porter and remove stopwordsfor all documents. Then, the Indexation task is performed using the *Lemur*[1]software.Each document is represented, following the model of Salton [16], as a weight vector $d_i^T = (w_{i,1}, \ldots, w_{i,j}, \ldots w_{i,|T|})$ where $w_{i,j}$ represents the weight of the term $t_j$ in a document$d_i$. This weight is calculated as the product of two factors $tf_{i,j}$and$idf_j$.$tf_{i,j}$factor corresponds to the frequency of appearance of the term $t_j$ in the document $d_i$ and $idf_j$ factor measuring the inverse of the frequency of the word in the corpus. Thus, the weight$w_{i,j}$is even higher than the term $t_j$ is frequent in document $d_i$and scarce throughout the corpus.

Factors for the calculation of *tf* and *idf*, we use the formulations defined by Robertson [17]:

$$tf_{i,j} = \frac{k_1 n_{i,j}}{n_{i,j} + k_2(1 - b + b\frac{|d_i|}{d_{avg}})}$$

where$n_{i,j}$ is the number of occurrences of term$t_j$ in document$d_i$, $|d_i|$is the document size $d_i$and $d_{avg}$ is the average size of all documents in the corpus.$k_1$, $k_2$and b are three constants that are the respective values 1, 1 and 0.5.

$$idf_j = log\frac{|D| - \left|\{d_i|t_j \in d_i\}\right| + 0.5}{\left|\{d_i|t_j \in d_i\}\right| + 0.5}$$

where$|D|$ is the size of the corpus and $\left|\{d_i|t_j \in d_i\}\right|$ is the number of documents in the corpus where the term tj appears at least once.

### 3.2.Visual indexing of medical reports

The representation of the visualmodality in medical reports is performed in two phases: the creation of a visual vocabulary and medical image representation using this vocabulary.

The vocabulary V of the visual modality is obtained using the approach "*bag-of-words*" [18]. The process involves three steps: the selection of regions or points of interest, the descriptionthrough the calculation of the points or regions descriptor and the combination of descriptors classes constituting the visual words.

We use two different approaches for the first two steps.

The first step uses the characterization of images with regions of interest detected by the Maximally Stable Extremal Regions (MSER) [19] and represented by their bounding ellipses (using the method proposed by [20]). These regions are then described by the descriptor Scale Invariant Feature Transform (SIFT)[26].

For the second step, classclusteris performed by applying the k-means algorithm on the set of descriptors to obtain k clusters descriptors. Each cluster center presents then a visual word.

---

[1]http://www.lemurproject.com

The representation of an image using the previously vocabulary defined to calculate a weight vector $d_i^v$ exactly as the textual modality. For visual words of the image, we first calculate the descriptors of points or regions of the image, and then associate with each descriptor the word vocabulary nearest to the sense of the *Euclidean distance*.

## Combination of modalities with latent semantic

In this section, we show how the two vocabularies are combined using the technique of latent semantic. A Latent Semantic Indexing (LSI) was first introduced in the field of information retrieval by [21, 22]. This technique is to reduce the matrix indexing in new space sensible dimensions expressed more "*semantic*". This reduction is intended to show the hidden semantic relationships in the co-occurrence. This is called the latent semantic. This latent semantic allows for example to reduce the effects of *Synonymy* and *Polysemy*. It is also used to index without translation or dictionary, parallel corpora, ie composed of documents in different languages, but supposed to be translate of each other.

Technically, the LSA method is a manufacturing process of the matrix $M$ co-occurrence between terms and documents. Indeed, it is in fact the *singular value decomposition* (SVD) of the matrix $M$: $M_{i,j}$ describes the occurrences of term $i$ in document $j$. The goal is to compute the matrices $U$, $\Sigma$ and $V$ such that:

$$M = U\Sigma V^t$$

Where

- U is the matrix of eigenvectors of $MM^t$
- $V^t$ is matrix of eigenvectors of $M^tM$
- $\Sigma$ is the diagonal matrix of singular values $r \ x \ r$
-

This transformation allows us to represent the matrix $M$ as a product of two different sources of information: matrix $U$ to documents and the second matrix M relative to the terms. Using the k largest eigenvalues of $\Sigma$ and truncating the matrices $U$ and $V$ accordingly, we obtain an approximation of rank $k$ of $M$:

$$M_k = U_k\Sigma_k V_k^t$$

where $k < r$ is the dimension of the space latent. This reduction in size allows you to capture important information and eliminate noise information produced by the redundant information, such as synonymy and *Polysemy*. It should be noted that the choice of the parameter k is difficult because it must be large enough to not lose information, and small enough to play its role of redundancy reduction.

## 4. AUTOMATIC MEDICAL IMAGE ANNOTATION

The automatic medical image annotation is quite similar to the indexing described in the previous section. As input, we have an un-annotated medical image. However, the automatic medical image annotation is to find a correlation between textual and visual information to be able to automatically assign a textual annotation of new images. To do this, we need to calculate the joint probability between words and visual terms. We have a similar approach with the method described in [23]. Figure 4 illustrates the process of medical image annotation.

For medical image un-annotated vector q is constructed as we have done in Section 3.2. We find, then, the similar images to the query image using the vector model with a similarity measure more two representations contain the same elements, the more likely they represent the same information are high. The calculation of the relevance value of an image to the query *RSV* (Retrieval Status Value) is calculated, based on the similarity of query vector, using the cosine function and we sort the results in descending order and we retain the image that has the highest score. We attribute this new image non-annotated terms that are associated with the image that has the best score.

$$RSV(\vec{Q}, \vec{D}) = \frac{\vec{Q} \cdot \vec{D}}{|\vec{Q}| X |\vec{D}|} = \frac{\sum w_{iQ} X w_{iD}}{\sqrt{\sum w_{iQ}^2} X \sqrt{\sum w_{iD}^2}}$$

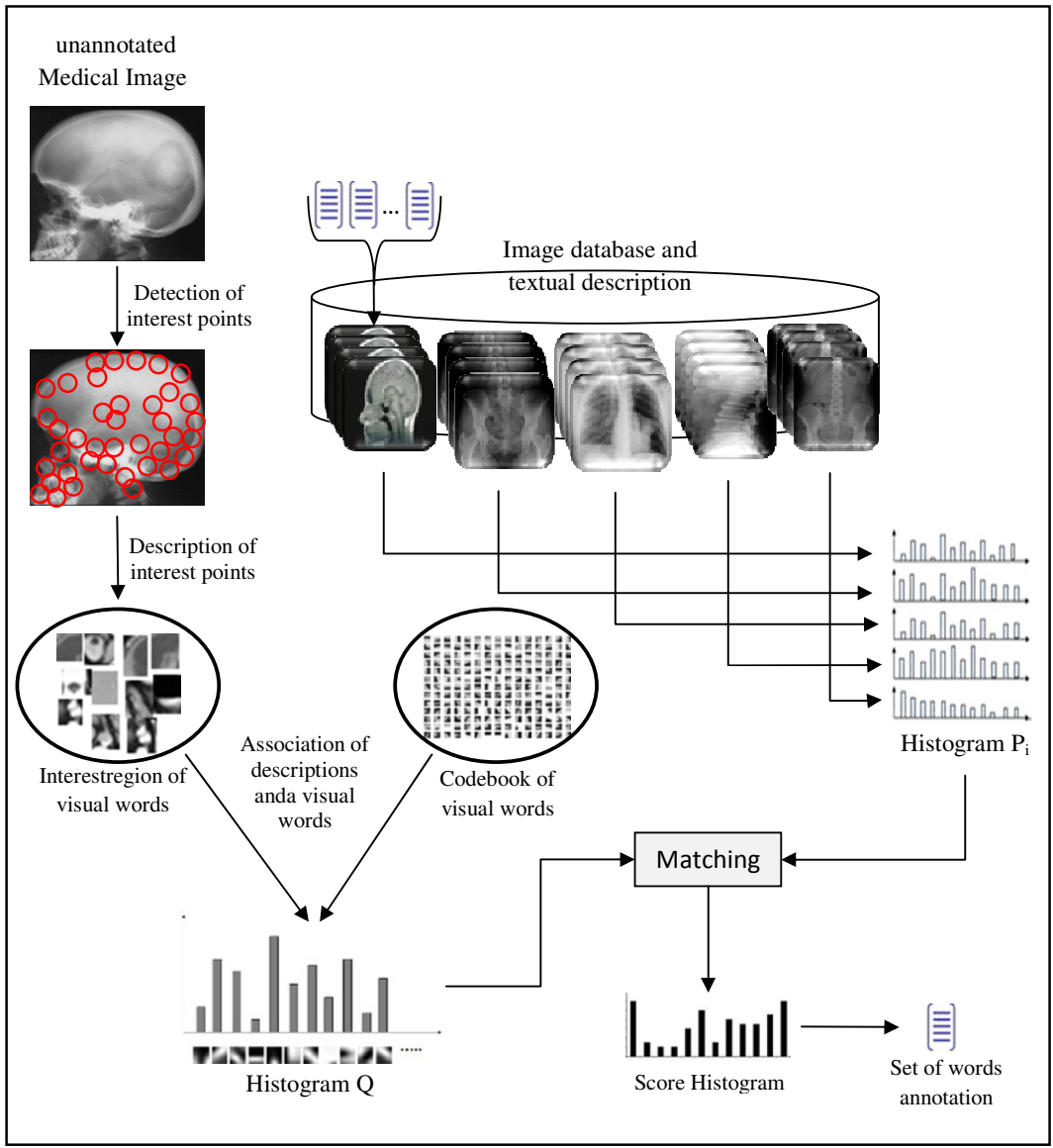

Figure4.Automatic medical image annotation

## 5. EXPERIMENTAL EVALUATION

### 5.1. Data test and evaluation criteria

The relevance of our approach is evaluated on the annotation collection provided by the military Tunisian hospital. This collection consists than 1000 medical reports on radiology images classified in five categories (*thorax*, *abdomen*, *lumbar spine*, *pelvis* and *skull*). Each report consists of a radiological image and text. We treated five types of reports for each type of radiological images; Figure 5 is an example of radiological images processed in this experiment. We started with a pre-indexing phase reports using the proposed approach which combines text indexing using the technique of *tf.idf* with indexing using the technique of visual bag of visual word. To determine the co-occurrence between the textual and visual descriptors and also to avoidethe semantic gap between these two, we implement the matrix obtained latent semantics. At this level, we have a co-occurrence between textual and visual words,, which makes labeling regions of radiological images with textual words.

To evaluate our system for automatic annotation of a radiological image. We use the criterion of average precision (*Mean Average Precision* - MAP), which is a standard criterion in information retrieval, is then used to evaluate the relevance of the results.

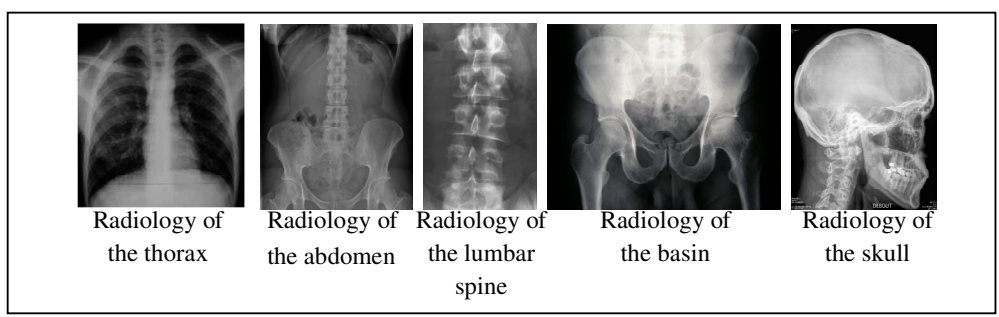

| Radiology of the thorax | Radiology of the abdomen | Radiology of the lumbar spine | Radiology of the basin | Radiology of the skull |

Figure5.Examples of radiological images training.

### 5.2. Results

A preliminary step of our experiment is to index the medical reports using the model we have proposed that combines between the textual description and visual description and reduced by latent semantic.

In fact, to demonstrate the contribution of our approach using automatic annotation of medical images, we present Table 1, which summarizes the obtained valuesfor each MAP testing. We note that our approach works well on images of radiology of the skull compared to other types of images. These global observations are confirmed by the curves of precision / recall shown in Figure 2.

| Type of radiographic image | MAP |
|---|---|
| Thorax | 0.1057 |
| Abdomen | 0.1326 |
| Lumbar spine | 0.2554 |
| Basin | 0.2826 |
| Skull | 0.3359 |

Table 1.Average precision results obtained for different types of radiographic images.

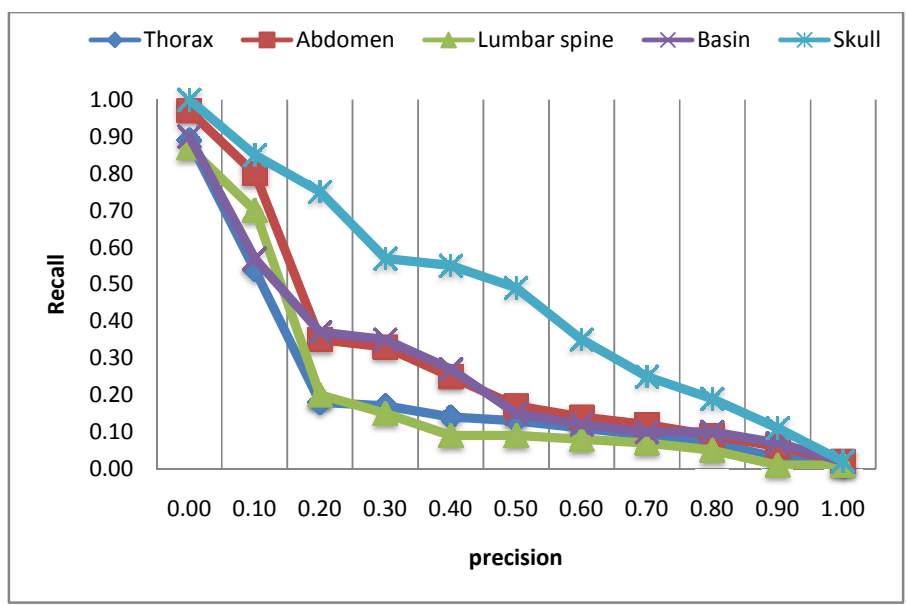

Figure 6.Courbe precision / recall for different types of radiological images processed.

## 6. CONCLUSION

We present in this paper an approach to medical image annotation based primarily on a method of indexing medical reports, including both information extracted will be represented using an approach "*bag-of-words*" and by latent semantic factorize. The main contribution of our system is that it uses a thesaurus medical to extract concepts and minimize the maximum indexing field. For the task of automatic annotation of medical images, the LSA technique significantly improves the quality of the annotation by combining different visual feature (i.e. local descriptor and global feature). In addition, the LSA is used to reduce the complexity computations on large matrices across maintaining good performance. Finally, we conducted an experiment made on five types of radiological imaging to evaluate the performance of our system of medical annotation. The results showed that our approach works better with more images from the radiology of the skull.